\documentclass{elsarticle}

\usepackage{lineno}
\usepackage{amsmath}
\usepackage{amssymb}
\usepackage{color}
\modulolinenumbers[1]

\journal{Aeolian Research}

\biboptions{authoryear}

\begin{document}

\begin{frontmatter}

\title{Aeolian sand transport: Scaling of mean saltation length and height and implications for mass flux scaling}

\author{Thomas P\"ahtz}
\address{Institute of Port, Coastal and Offshore Engineering, Ocean College, Zhejiang University, 866 Yu Hang Tang Road, 310058 Hangzhou, China}

\author{Katharina Tholen}
\address{Institute for Theoretical Physics, Leipzig University, Br\"uderstra{\ss}e 16, 04103 Leipzig, Germany}

\begin{abstract}
Wind tunnel measurements of the mean saltation length $L$ and of different proxies of the mean saltation height $H$ in saturated aeolian sand transport indicate that $L$ and $H$ are relatively insensitive to both the wind speed and grain diameter $d$. The latter result is currently unexplained and contradicts the theoretical prediction $L\propto H\propto d$. This prediction is based on the assumption that the characteristic velocity $\sqrt{\tilde gd}$ of bed grains ejected by the splash of an impacting grain controls the average saltation kinematics. Here, we show that a recent analytical saltation model that considers only rebounds of saltating grains, but neglects splash ejection, is consistent with the measurements. The model suggests that the buffer layer of the inner turbulent boundary layer, which connects the viscous sublayer with the log-layer, is partially responsible for the insensitivity of $L$ and $H$ to $d$. In combination, the measurements and model therefore indicate that splash ejection, though important to sustain saltation, does not significantly affect the average saltation kinematics. This finding represents a strong argument against the \cite{UngarHaff87}-scaling and in favor of the \cite{Duranetal11}-scaling of the saturated saltation mass flux, with implications for ripple formation on Mars. Furthermore, it supports the recent controversial claim that this flux is insensitive to soil cohesion.
\end{abstract}

\begin{keyword}
aeolian sand transport \sep saltation trajectories \sep mass flux \sep ripple formation \sep cohesion
\end{keyword}

\end{frontmatter}

%\linenumbers

\section{Introduction}
The evolution of windblown dry planetary sand surfaces is determined by spatial changes of the mass flux $Q$ associated with sand transport. Since the seminal work by \cite{Bagnold41}, there has been an ongoing debate about the scaling of $Q$ for saturated (i.e., equilibrium) transport conditions. A large part of this debate has focused on the question of how $Q$ increases with the wind friction velocity $u_\ast$. While many theoretical studies proposed various cubic relationships between $Q$ on $u_\ast$ \citep[e.g.,][]{Bagnold41,Owen64,Sorensen04,DuranHerrmann06}, a consensus has emerged in the last two decades in favor of the quadratic relationship $Q\sim u_\ast^2-u_{\ast t}^2$ \citep[e.g.,][]{UngarHaff87,Duranetal11,Koketal12,Valanceetal15}, where $u_{\ast t}$ is the dynamic threshold value of $u_\ast$ below which $Q$ is predicted to vanish. This quadratic relationship has been directly confirmed by sophisticated wind tunnel \citep{Creysselsetal09,Hoetal11} and field measurements of $Q(u_\ast)$ \citep{Martinetal13,MartinKok17}. Additionally, it is supported by wind tunnel \citep{RasmussenSorensen08,Creysselsetal09,Hoetal11,Hoetal14} and field measurements \citep{Greeleyetal96,Namikas03,MartinKok17} showing nearly $u_\ast$-independent average kinematic transport layer properties. Recently, it has been predicted \citep{PahtzDuran20} and subsequently experimentally confirmed \citep{Ralaiarisoaetal20} that the quadratic scaling turns into a quartic one, $Q\sim u_\ast^4$, for very intense aeolian sand transport. The reason is a transition from a saltation regime, in which transported grains bounce along the surface in ballistic trajectories, to a collisional regime, in which grain trajectories are substantially affected by midair collisions \citep{Carneiroetal13,PahtzDuran20,Ralaiarisoaetal20}.

However, the debate about the scaling of $Q$ has not been settled. While there is a consensus in favor of $Q\sim u_\ast^2-u_{\ast t}^2$ for the saturated saltation regime, different research groups have made two different predictions about the scaling of the prefactors in this relationship:
\begin{linenomath}
\begin{alignat}{2}
 \text{\cite{UngarHaff87}:}\quad&&Q&\propto\frac{\rho_f\sqrt{\tilde gd}}{\tilde g}(u_\ast^2-u_{\ast t}^2), \label{UH87} \\
 \text{\cite{Duranetal11}:}\quad&&Q&\propto\frac{\rho_fu_{\ast t}}{\tilde g}(u_\ast^2-u_{\ast t}^2), \label{D11}
\end{alignat}
\end{linenomath}
where $\rho_f$ is the fluid density, $d$ is the median grain diameter, and $\tilde g\equiv(1-\rho_f/\rho_p)g$, with $\rho_p$ the particle density, is the buoyancy-reduced value of the gravitational constant $g$. Equation~(\ref{UH87}), first proposed by \cite{UngarHaff87}, is favored by \cite{JenkinsValance14}, \cite{Valanceetal15}, \cite{Berzietal16}, and \cite{Ralaiarisoaetal20}, while Eq.~(\ref{D11}) is favored by \cite{Duranetal11}, \cite{Koketal12}, \cite{PahtzDuran20}, and \cite{Pahtzetal21}.

Eqs.~(\ref{UH87}) and (\ref{D11}) are based on two different physical understandings of the so-called \textit{replacement capacity} condition, that is, the bed boundary condition that ensures that, on average, a single grain leaves the sand bed per grain impacting the bed. Proponents of Eq.~(\ref{UH87}) argue that both the rebound of the impacting grain and the ejection of new bed grains should be considered in the modeling of the replacement capacity condition \citep{Valanceetal15}, which implies that $\sqrt{\tilde gd}$ is the predominant velocity scale controlling the average saltation kinematics \citep{Berzietal16}. In contrast, proponents of Eq.~(\ref{D11}) argue that only rebounding grains should be considered \citep{Pahtzetal20a}, which implies that the threshold wind friction velocity $u_{\ast t}$ is the predominant velocity scale \citep{PahtzDuran20}. These two physical understandings are mutually exclusive, since it is agreed upon that $u_{\ast t}/\sqrt{\tilde gd}$ is generally not a constant. Rather, it depends on further dimensionless quantities, such as the particle-fluid-density ratio $s\equiv\rho_p/\rho_f$ and Galileo number $Ga\equiv d\sqrt{s\tilde gd}/\nu$, where $\nu$ is the kinematic fluid viscosity \citep{Kok10b,Berzietal16,Berzietal17,PahtzDuran18a,Pahtzetal20a,Andreottietal21,Pahtzetal21}. That is, the predictions of Eqs.~(\ref{UH87}) and (\ref{D11}) can be very different from each other across atmospheric conditions.

\cite{Hoetal14} carried out wind tunnel experiments of saturated aeolian sand transport for various $u_\ast$ and two sands with significantly different grain diameters ($d=[232,630]~\mathrm{\mu m}$). They measured the mean saltation length $L$, the mean saltation height $H_{10d}$ of grains hopping at least $10d$ high, and the so-called \textit{focus height} $H_f$ (i.e., the highest elevation below which the local wind velocity is insensitive to $u_\ast$), the latter two of which are proxies for the mean saltation height $H$ of all grains. Importantly, they noted that the corresponding measured values $L\approx[60,75]~\mathrm{mm}$ and $H_{10}\approx[5.1,7.8]~\mathrm{mm}$ (insensitive to $u_\ast$) and $H_f\approx[8.3,12.6]~\mathrm{mm}$ are inconsistent with the notion that the average saltation kinematics are controlled by the velocity scale $\sqrt{\tilde gd}$ as this notion would imply $L\propto H\propto d$. Similarly, from the surface roughness data of \cite{Rasmussenetal96} for five different sands within the range $d\in[125,544]~\mathrm{\mu m}$, \cite{Andreotti04} (his Fig.~12(b)) estimated that $H_f$ scales with $\sqrt{d}$ rather than with $d$ (see also inset of Fig.~24 of \cite{Duranetal11}). Both data sets cause doubt on Eq.~(\ref{UH87}) and remain unexplained.

Here, we show that the recent analytical saltation model of \cite{Pahtzetal21} (introduced in section~\ref{AnalyticalModel}), which is based on the same physical understanding of the replacement capacity condition as Eq.~(\ref{D11}), is consistent with both data sets (section~\ref{Results}). This model suggests that $L$ and $H$ obey a relatively simple previously unreported scaling behavior.

\section{Analytical Saltation Model of \cite{Pahtzetal21}} \label{AnalyticalModel}
The analytical saltation model of \cite{Pahtzetal21} predicts the average saltation kinematics for transport threshold conditions ($u_\ast=u_{\ast t}$). We assume that this model also applies to saltation conditions far from the threshold because experiments indicate that the average saltation kinematics is insensitive to $u_\ast$ \citep{Hoetal14}. Since only a few grains are in motion near threshold conditions, the model assumes that saltation is driven by an undisturbed inner turbulent boundary layer velocity profile $u_x$, which is a function of $u_\ast$, the particle Reynolds number $Re_d\equiv u_\ast d/\nu$, and the nondimensionalized elevation above the sand bed surface $z/d$:
\begin{linenomath*}
\begin{equation}
 u_x=u_\ast f_u(Re_d,z/d).
\end{equation}
\end{linenomath*}
For $Re_d\lesssim70$, the function $f_u$ covers two extreme regimes \citep{Julien10}:
\begin{linenomath*}
\begin{subequations}
\begin{alignat}{2}
 \text{Viscous sublayer ($Re_dz/d\lesssim5$):}\quad&&f_u(Re_d,z/d)&=Re_dz/d, \label{ViscousProfile} \\
 \text{Log-layer ($Re_dz/d\gtrsim30$):}\quad&&f_u(Re_d,z/d)&=\kappa^{-1}\ln(z/z_o), \label{LogProfile}
\end{alignat}
\end{subequations}
\end{linenomath*}
where $\kappa=0.4$ is the von K\'arm\'an constant and \citep{GuoJulien07,Sleath84}
\begin{linenomath*}
\begin{equation}
 z_o=\frac{d}{9Re_d}+\frac{d}{30}\left[1-\exp\left(-\frac{Re_d}{26}\right)\right]
\end{equation}
\end{linenomath*}
is the surface roughness across smooth ($Re_d\lesssim4$) and rough ($Re_d\gtrsim70$) conditions. The buffer layer ($5\lesssim Re_dz/d\lesssim30$) connects the viscous sublayer with the log-layer. However, for $Re_d\gtrsim70$, only the log-layer exists. An empirical unifying expression across all regimes, the so-called \textit{Law of the Wall}, is given by \citep{GuoJulien07,Julien10}
\begin{linenomath*}
\begin{equation}
\begin{split}
 &f_u(Re_d,z/d)=7\arctan\left(\frac{Re_dz/d}{7}\right)+\frac{7}{3}\arctan^3\left(\frac{Re_dz/d}{7}\right) \\
 &-0.52\arctan^4\left(\frac{Re_dz/d}{7}\right)+\ln\left[1+\left(\frac{Re_dz/d}{B_\kappa}\right)^{(1/\kappa)}\right] \\
 &-\frac{1}{\kappa}\ln\left\{1+0.3Re_d\left[1-\exp\left(-\frac{Re_d}{26}\right)\right]\right\}, \label{uxcomplex}
\end{split}
\end{equation}
\end{linenomath*}
where $B_\kappa=\exp(16.873\kappa-\ln9)$. Note that there are alternative empirical expressions that work similarly well \citep{Julien10}. However, to be consistent with \cite{Pahtzetal21}, Eq.~(\ref{uxcomplex}) is used in the model unless mentioned otherwise.

The model represents the entire grain motion by grains saltating in identical periodic trajectories. That is, it combines the equations of motion for an individual saltating grain, linking the grain impact velocity $\mathbf{v_\downarrow}$ to the grain lift-off (or rebound) velocity $\mathbf{v_\uparrow}$ (i.e., $\mathbf{v_\downarrow}=f_1(\mathbf{v_\uparrow})$), with boundary conditions that describe the average outcome of a grain-bed rebound ($\mathbf{v_\uparrow}=f_2(\mathbf{v_\downarrow})$). For given values of $Ga$, $s$, and $v_{\uparrow z}$, the model then calculates the value of the Shields number $\Theta\equiv u_\ast^2/(s\tilde gd)$ for which the grain trajectory is periodic (i.e., $f^{-1}_2=f_1$), summarized by the following set of equations (explained shortly):
\begin{linenomath*}
\begin{subequations}
\begin{align}
 \hat v_{\downarrow z}&=-1-W\left[-(1+\hat v_{\uparrow z})e^{-(1+\hat v_{\uparrow z})}\right], \label{vdown} \\
 |\mathbf{\hat v_\uparrow}|/|\mathbf{\hat v_\downarrow}|&=A+B\hat v_{\downarrow z}/|\mathbf{\hat v_\downarrow}| \label{e} \\
 -\hat v_{\uparrow z}/\hat v_{\downarrow z}&=A/\sqrt{-\hat v_{\downarrow z}/|\mathbf{\hat v_\downarrow}|}-B, \label{ez} \\
 \mu_b(1+\hat v_{\uparrow z})+\hat v_{\uparrow x}&=\frac{\sqrt{\Theta}}{v_{s\ast}}f_u\left\{Ga\sqrt{\Theta},sv_{s\ast}^2[-\hat v_{\downarrow z}(1+\hat v_{\uparrow z})-\hat v_{\uparrow z}]+Z_\Delta\right\}, \label{Theta}
\end{align}
\end{subequations}
\end{linenomath*}
where the velocities have been nondimensionalized via $\mathbf{\hat v_{\uparrow(\downarrow)}}\equiv\mathbf{v_{\uparrow(\downarrow)}}/v_s$, with $v_s$ the average terminal settling velocity of saltating grains, calculated by
\begin{linenomath*}
\begin{equation}
\begin{split}
 v_{s\ast}\equiv\frac{v_s}{\sqrt{s\tilde gd}}&=\frac{1}{\mu_b}\left[\sqrt{\frac{1}{4}\sqrt[m]{\left(\frac{24}{C_d^\infty Ga}\right)^2}+\sqrt[m]{\frac{4\mu_b}{3C_d^\infty}}}-\frac{1}{2}\sqrt[m]{\frac{24}{C_d^\infty Ga}}\right]^m, \label{Settling} \\
 \text{with}\quad\mu_b&\equiv\frac{\hat v_{\downarrow x}-\hat v_{\uparrow x}}{\hat v_{\uparrow z}-\hat v_{\downarrow z}}.
\end{split}
\end{equation}
\end{linenomath*}
Eqs.~(\ref{vdown})-(\ref{Theta}) and (\ref{Settling}) were derived starting from linearizing the fluid drag acceleration acting on saltating grains via approximating the fluid-particle-velocity difference as $|\mathbf{u}-\mathbf{v}|\approx\overline{u_x}-\overline{v_x}$, where the overbar represents the saltation trajectory average. This linearization then led to a decoupling of the equation of vertical motion from the equation of streamwise motion, allowing them to be readily solved analytically. Eq.~(\ref{vdown}), in which $W$ denotes the principal branch of the Lambert-$W$ function, was derived from the exact solution of this decoupled equation of vertical motion. Eqs.~(\ref{e}) and (\ref{ez}), where $A=0.87$ and $B=0.72$, are empirical laws describing the mean rebound restitution coefficient $e$ and vertical rebound restitution coefficient $e_z$ in agreement with measurements by \cite{Beladjineetal07}, who carried out experiments of the collision process between an incident grain and a three-dimensional granular packing. Eq.~(\ref{Theta}), where $Z_\Delta=0.7$ is the dimensionless elevation of a grain's center of mass during its rebound with the sand bed relative to the virtual zero-level of $u_x$, was derived from an approximate solution of the equation of streamwise motion. Eq.~(\ref{Settling}) is a direct consequence of the linearization of the fluid drag law by \cite{Camenen07}, where the values of the parameters $m$ and $C_d^\infty$ depend on the grain shape (discussed later).

For given values of $Ga$, $s$, and $\hat v_{\uparrow z}$, solving Eqs.~(\ref{vdown})-(\ref{Theta}) and (\ref{Settling}) yields a unique value of $\Theta$. The saltation threshold Shields number $\Theta_t\equiv u_{\ast t}^2/(s\tilde gd)$ is then calculated as the smallest value of $\Theta$ for which a solution exists:
\begin{linenomath*}
\begin{equation}
 \Theta_t(Ga,s)=\min_{\hat v_{\uparrow z}}\Theta(Ga,s,\hat v_{\uparrow z}).
\end{equation}
\end{linenomath*}
From the value of $\hat v_{\uparrow z}$ corresponding to this threshold solution, we then obtain $L$ and $H$ via Eqs.~(\ref{vdown}) and (\ref{Settling}) and
\begin{linenomath*}
\begin{align}
 \frac{L}{sdv_{s\ast}^2}&=(\hat v_{\uparrow z}-\hat v_{\downarrow z})\left\{\frac{\sqrt{\Theta_t}}{v_{s\ast}}f_u\left[Ga\sqrt{\Theta_t},\frac{1}{2}sv_{s\ast}^2(\hat v_{\uparrow z}+\hat v_{\downarrow z})+Z_\Delta\right]-\mu_b\right\}, \label{L} \\
 \frac{H}{sdv_{s\ast}^2}&=\hat v_{\uparrow z}-\ln(1-\hat v_{\uparrow z}). \label{H}
\end{align}
\end{linenomath*}
Like Eqs.~(\ref{vdown})-(\ref{Theta}) and (\ref{Settling}), Eqs.~(\ref{L}) and (\ref{H}) were derived from the equation of streamwise and vertical motion, respectively, associated with the linearized fluid drag law.

\section{Results} \label{Results}
\begin{figure}[htb!]
 \begin{center}
  \includegraphics[width=1.0\columnwidth]{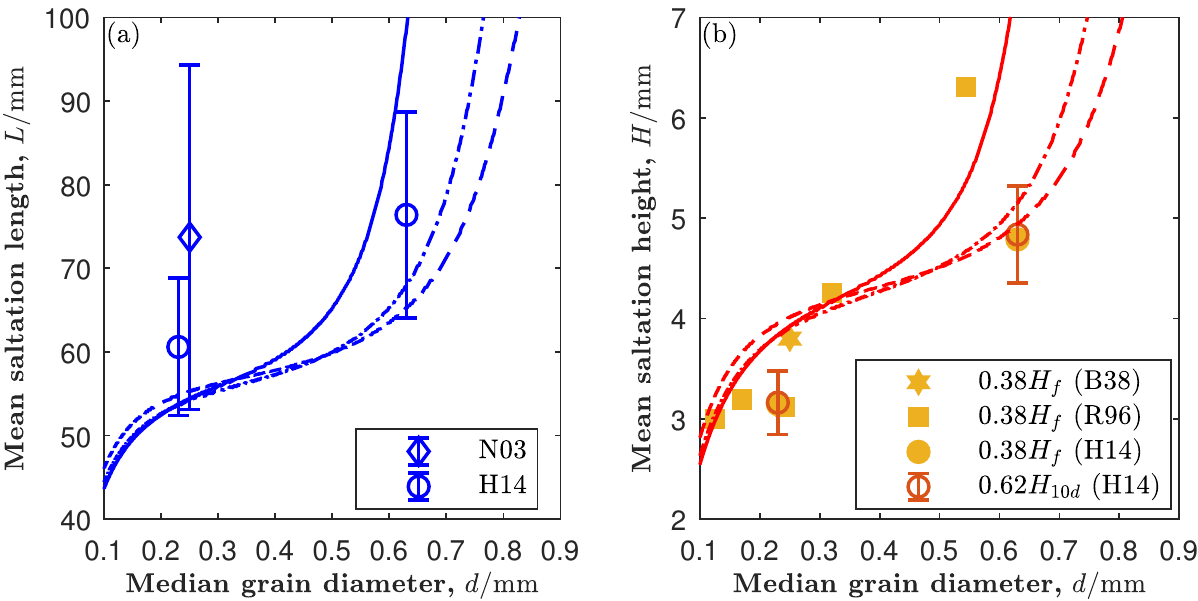}
 \end{center}
 \caption{(a) Mean saltation length $L$ and (b) mean saltation height $H$ versus median grain diameter $d$. Symbols correspond to experimental data of $L$ and two different proxies of $H$: the mean saltation height $H_{10d}$ of grains hopping at least $10d$ high and the focus height $H_f$ \citep{Bagnold38,Rasmussenetal96,Namikas03,Hoetal14}. Lines correspond to model predictions using different parameter values of $m$ and $C^\infty_d$, corresponding to different grain shapes (see text). The environmental parameters used for the model predictions are the same as in \cite{Hoetal14}: $\rho_p=2470~\mathrm{kg/m^3}$, $\rho_f=1.2~\mathrm{kg/m}^3$, and $\nu=1.6\times10^{-5}~\mathrm{m^2/s}$.}
\label{ModelvsMeasurements}
\end{figure}

Figure~\ref{ModelvsMeasurements} compares the model predictions of the mean saltation length $L$ and height $H$ with measurements. The comparison includes wind tunnel measurements of $L$ and $H_{10d}$ by \cite{Hoetal14} (symbols labeled `H14'), who used a novel measurement method that, in contrast to previous methods, did not require tracking the trajectories of single grains. \cite{Hoetal14} also reported field data of $L$ extracted from applying their method to the streamwise mass flux distribution field data by \cite{Namikas03} (symbols labeled `N03'). Note that data obtained from older grain tracking-based methods \citep[e.g.,][]{Nalpanisetal93,Zhangetal07} are prone to error for various reasons, such as the fact that the grains' lateral motion may cause them to leave the observation window \citep{Hoetal14}, and are therefore excluded from the comparison. The comparison in Fig.~\ref{ModelvsMeasurements} also includes direct wind tunnel measurements of the focus height $H_f$ by \cite{Hoetal14}, field measurements of $H_f$ by \cite{Bagnold38} (symbols labeled `B38'), and estimations of $H_f$ by \cite{Andreotti04} (see his Fig.~12(b) or the inset of Fig.~24 of \cite{Duranetal11}), which he obtained from the surface roughness wind tunnel data of \cite{Rasmussenetal96} (symbols labeled `R96'). Note that \cite{Hoetal14} reported a further field measurement of $H_f$ for the experiments by \cite{Namikas03}, which they estimated from the velocity profile data reported in \cite{Namikasetal03}. However, we were unable to reproduce their estimation, since those velocity profile data in \cite{Namikasetal03} that correspond to the experiments by \cite{Namikas03} exhibit a too narrow range of $u_\ast$ (varying only between $0.292~\mathrm{m/s}$ and $0.387~\mathrm{m/s}$) to reliable estimate $H_f$.

Because both $H_{10d}$ and $H_f$ are proxies of $H$ and proportional to each other ($H_{10d}\simeq1.63H_f$) for Earth's atmospheric conditions \citep{Hoetal14}, we have assumed $H\propto H_{10d}\propto H_f$ and fitted the proportionality constant $H_f/H$ to best agreement between model predictions and measurements of $H$, yielding $H_f/H=0.38$ and thus $H_{10d}/H=0.62$. However, we note that these proportionalities may not necessarily hold generally across atmospheric conditions.

The three different model predictions in Fig.~\ref{ModelvsMeasurements} correspond to three different combinations of $m$ and $C^\infty_d$. The values $m=2.0$ and $C_d^\infty=0.4$ (solid lines) are approximately valid for spherical grains, while smaller values of $m$ and larger values of $C_d^\infty$ are associated with more angular grains \citep{Camenen07}. In particular, the values $m=1.5$ and $C^\infty_d=1.0$ (dash-dotted lines) are close to those recommended for natural sand grains by \cite{Camenen07} and the values $m=1.0$ and $C^\infty_d=1.5$ (dashed lines) are those that \cite{Hoetal14} used to reproduce their measured saltation length and height distributions with their numerical model. Consistently, it can be seen in Fig.~\ref{ModelvsMeasurements} that the model prediction corresponding to $m=1.0$ and $C^\infty_d=1.5$ agrees better with the largest-$d$ data points of \cite{Hoetal14} than the model prediction corresponding to $m=2.0$ and $C_d^\infty=0.4$, whereas the opposite is true for the largest-$d$ data point of \cite{Rasmussenetal96}. In contrast, for sufficiently small $d$, the different model predictions are all consistent with the experimental data and differ only very slightly from each other. This is because the drag force is dominated by its Stokes drag component for sufficiently small $d$, resulting in the dimensionless settling velocity $v_{s\ast}\simeq Ga/18$ (Eq.~(\ref{Settling}) for small Galileo number $Ga$), which is independent of $m$ and $C_d^\infty$. Note that it is not surprising that different experimental data sets require different combinations of $m$ and $C_d^\infty$ for good model agreement, since typical shapes of sand grains can vary substantially across sites on Earth \citep{Raffaeleetal20}.

\begin{figure}[htb!]
 \begin{center}
  \includegraphics[width=1.0\columnwidth]{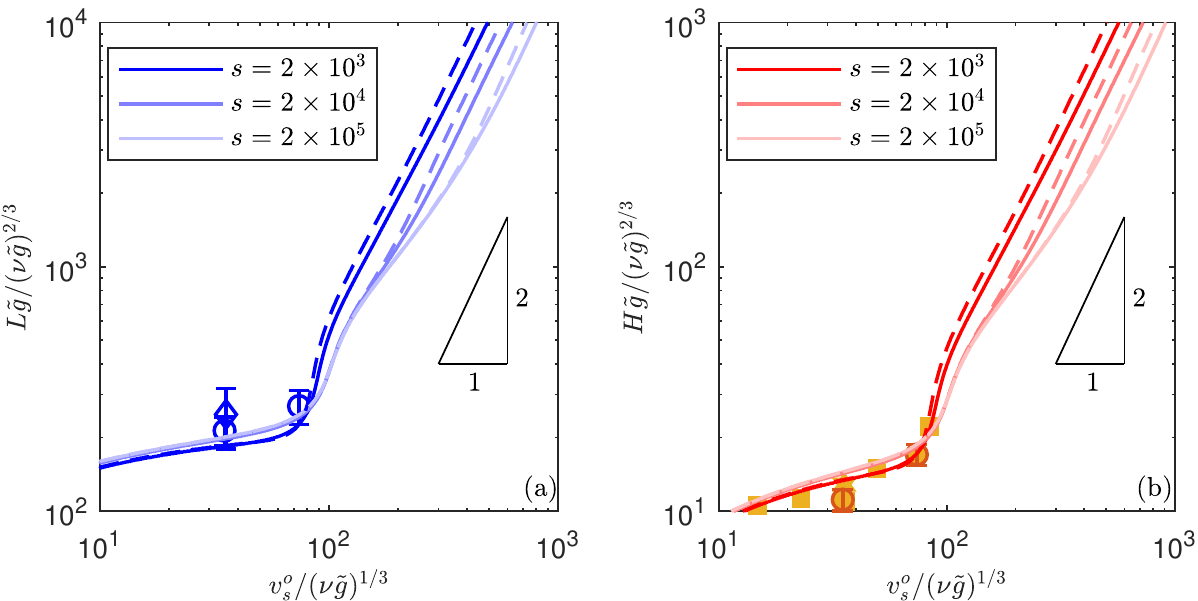}
 \end{center}
 \caption{(a) Nondimensionalized mean saltation length $L\tilde g/(\nu\tilde g)^{2/3}$ and (b) nondimensionalized mean saltation height $H\tilde g/(\nu\tilde g)^{2/3}$ versus nondimensionalized settling velocity $v^o_s/(\nu\tilde g)^{1/3}$. Symbols correspond to experimental data of $L$ and two different proxies of $H$: the mean saltation height $H_{10d}$ of grains hopping at least $10d$ high and the focus height $H_f$ \citep{Bagnold38,Rasmussenetal96,Namikas03,Hoetal14}. For symbol legend, see Fig.~\ref{ModelvsMeasurements}. For the experimental data by \cite{Bagnold38}, \cite{Rasmussenetal96}, and \cite{Namikas03}, the parameter values $m=2.0$ and $C^\infty_d=0.4$ are used to calculate $v^o_s$ via Eq.~(\ref{Settling}), whereas for the experimental data by \cite{Hoetal14}, the parameter values $m=1.0$ and $C^\infty_d=1.5$ are used (cf. Fig.~\ref{ModelvsMeasurements}). Lines correspond to model predictions using different parameter values of $m$ and $C^\infty_d$ (same as the solid and dashed lines in Fig.~\ref{ModelvsMeasurements}), corresponding to different grain shapes (see text), and different values of the particle-fluid-density ratio $s$.}
\label{ModelvsMeasurementsDimensionless}
\end{figure}

In Fig.~\ref{ModelvsMeasurementsDimensionless}, we replotted the measurements and model predictions using dimensionless parameters that capture most of the variation with $m$ and $C^\infty_d$ and also most of the variation with the particle-fluid-density ratio $s$, namely, $L\tilde g/(\nu\tilde g)^{2/3}$, $H\tilde g/(\nu\tilde g)^{2/3}$, and $v^o_s/(\nu\tilde g)^{1/3}$, where $v^o_s\equiv v_s|_{\mu_b=\mu^o_b}$, with $\mu^o_b=\mu_b|_{v_{\downarrow z}=-v_{\uparrow z}}\simeq0.646$ \citep[i.e., $\mu_b$ in the limit of negligible vertical drag, obtained from Eqs.~(\ref{e}) and (\ref{ez}),][]{Pahtzetal21}. It can be seen that there are essentially two distinct regimes: one regime for $10\lesssim v^o_s/(\nu\tilde g)^{1/3}\lesssim80$ in which $L\tilde g/(\nu\tilde g)^{2/3}$ and $H\tilde g/(\nu\tilde g)^{2/3}$ are very slightly increasing (i.e., roughly constant) and another regime for $v^o_s/(\nu\tilde g)^{1/3}\gtrsim80$ in which roughly $L\propto H\propto v_s^{o2}/\tilde g$.

\begin{figure}[htb!]
 \begin{center}
  \includegraphics[width=0.5\columnwidth]{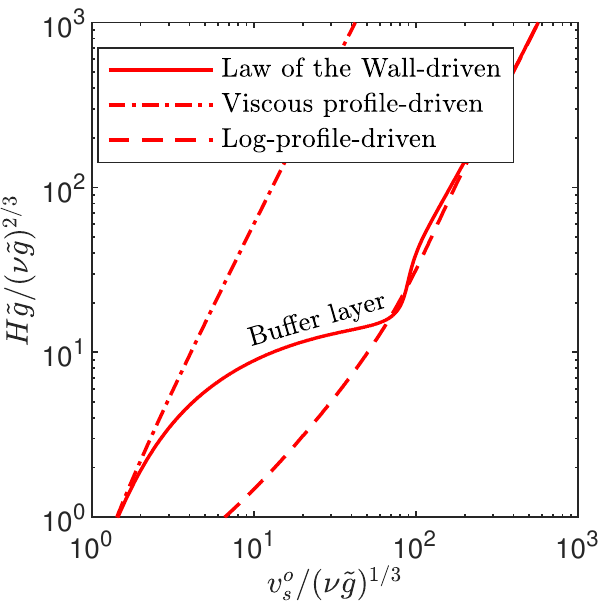}
 \end{center}
 \caption{Nondimensionalized mean saltation height $H\tilde g/(\nu\tilde g)^{2/3}$ versus nondimensionalized settling velocity $v^o_s/(\nu\tilde g)^{1/3}$ predicted by the model for a particle-fluid-density ratio $s=2000$ and three different driving flows: a general Law of the Wall driving flow (Eq.~(\ref{uxcomplex}), solid line), a pure viscous profile driving flow (Eq.~(\ref{ViscousProfile}), dash-dotted line), and a pure log-profile driving flow (Eq.~(\ref{LogProfile}), dashed line). The text ``Buffer layer'' indicates that the regime of roughly constant $H\tilde g/(\nu\tilde g)^{2/3}$ is strongly linked to the buffer layer of the inner turbulent boundary layer, which connects the viscous sublayer with the log-layer. The parameter values $m=2.0$ and $C^\infty_d=0.4$ are used to calculate the model predictions.}
\label{BufferLayer}
\end{figure}

The regime of roughly constant $L\tilde g/(\nu\tilde g)^{2/3}$ and $H\tilde g/(\nu\tilde g)^{2/3}$ is strongly linked to the buffer layer of inner turbulent boundary layer, which connects the viscous sublayer with the log-layer, as shown in Fig.~\ref{BufferLayer} for $H\tilde g/(\nu\tilde g)^{2/3}$. In fact, when replacing the general Law of the Wall driving flow (Eq.~(\ref{uxcomplex}), solid line in Fig.~\ref{BufferLayer}) in the model by a pure viscous profile driving flow (Eq.~(\ref{ViscousProfile}), dash-dotted line in Fig.~\ref{BufferLayer}) or a pure log-profile driving flow (Eq.~(\ref{LogProfile}), dashed line in Fig.~\ref{BufferLayer}), $H\tilde g/(\nu\tilde g)^{2/3}$ increases comparably strongly with $v^o_s/(\nu\tilde g)^{1/3}$.

\section{Discussion and Conclusions}
The analytical saltation model of \cite{Pahtzetal21} predicts the average saltation kinematics for transport threshold conditions. Using this model, assuming that the average saltation kinematics is insensitive to the wind friction velocity $u_\ast$ \citep{Hoetal14}, we have provided an explanation for previously unexplained measurements of a surprisingly weak dependency of the mean saltation length $L$ and height $H$ on the grain diameter $d$ (Fig.~\ref{ModelvsMeasurements}).

The model links this behavior to a regime in which $L\tilde g/(\nu\tilde g)^{2/3}$ and $H\tilde g/(\nu\tilde g)^{2/3}$ are roughly constant with the particle-fluid-density ratio $s$ and with the nondimensionalized settling velocity within the range $10\lesssim v^o_s/(\nu\tilde g)^{1/3}\lesssim80$ (Fig.~\ref{ModelvsMeasurementsDimensionless}). For $v^o_s/(\nu\tilde g)^{1/3}\gtrsim80$, the scaling changes to roughly $L\propto H\propto v_s^{o2}/\tilde g$, which is equivalent to $L\propto H\propto sd$ when the Galileo number $Ga$ is sufficiently large (from Eq.~(\ref{Settling})). Note that $v^o_s$, and thus this regime transition, depends on the typical shape of saltating grains. However, we are unaware of saltation trajectory wind tunnel or field data for large median grain diameters ($d\gtrsim700~\mathrm{\mu m}$ on Earth, see Fig.~\ref{ModelvsMeasurements}) that would be required to test this predicted regime transition. Note that we refrained from testing our model against measurements of the e-folding length of vertical concentration or mass flux profiles, which also exist for large $d$ \citep{DongQian07,Zhangetal17}, since measured values for similar conditions vary by more than an order of magnitude between the different studies \citep[][Table~5.3]{Ho12}, suggesting problems with some of the applied experimental methods. It is also unclear how strongly the saltation kinematics depends on grain size heterogeneity and turbulent wind fluctuations. The fact that the model, which assumes monodisperse sand and a mean turbulent flow, is consistent with field measurements (very heterogeneous sand and strong turbulent wind fluctuations), as shown in Figs.~\ref{ModelvsMeasurements} and \ref{ModelvsMeasurementsDimensionless} (see also Figs.~5(d) and 5(g) of \cite{Pahtzetal21}), hints on relatively weak effects. However, more data are needed to draw reliable conclusions.

By comparing model predictions for different driving flows, we have linked the regime of roughly constant $L\tilde g/(\nu\tilde g)^{2/3}$ and $H\tilde g/(\nu\tilde g)^{2/3}$ to the buffer layer of the inner turbulent boundary layer (Fig.~\ref{BufferLayer}), which connects the viscous sublayer with the log-layer. This leads to the conclusion that the viscous sublayer and buffer layer, which are usually assumed to be negligible \citep[e.g.,][]{Kok10b,Berzietal16,Berzietal17,LammelKroy17}, actually have a substantial impact on the statistical properties of saltation.

A further crucial difference between the model of \cite{Pahtzetal21} and previous analytical saltation models concerns the modeling of the replacement capacity condition, that is, the bed boundary condition that ensures that, on average, a single grain leaves the sand bed per grain impacting the bed. The difference lies in the notion of when a grain should be considered as leaving the bed. Most previous models assume that a grain can be considered as leaving the bed when it escapes the bed pocket in which it is resting, that is, when it is lifted a distance $\propto d$, which requires a lift-off velocity $\propto\sqrt{\tilde gd}$ \citep{ClaudinAndreotti06,Kok10b,Berzietal16,Berzietal17,Andreottietal21}. Under the approximation that grains impacting the bed nearly never fail to rebound, this assumption implies that $\sqrt{\tilde gd}$ is also the predominant velocity scale controlling the average saltation kinematics \citep{Berzietal16}. Under the further approximation that the vertical drag on saltating grains can be neglected, this implication leads to the prediction $L\propto H\propto\sqrt{\tilde gd}^2/\tilde g=d$. However, as noted in the introduction (see also Figs.~\ref{ModelvsMeasurements} and \ref{ModelvsMeasurementsDimensionless}), this prediction is inconsistent with measurements, suggesting potential problems with the modeling of the replacement capacity condition underlying this prediction.

One such potential problem was pointed out by \cite{PahtzDuran18a} and \cite{Pahtzetal20a}: Even if a grain initially escapes the bed surface pocket in which it is resting, it will still be quickly captured by the bed if it, on average, loses more energy in its subsequent impact(s) on the bed than it gains during its subsequent saltation hop(s) through wind drag acceleration (i.e., if the grain is net decelerated). In other words, grains should only be considered as leaving the bed if they are net accelerated after escaping their bed surface pocket \citep[note that a similar definition was recently also applied to aerodynamic entrainment,][]{JiaWang21}. The analytical saltation model of \cite{Pahtzetal21} therefore only considers rebounding grains in the modeling of the replacement capacity condition. That is, the velocity scale $\sqrt{\tilde gd}$ plays no role in this model, resulting in our novel predictions for $L$ and $H$. Consistently, the model predicts that, in the limit of threshold conditions, the minimal lift-off velocity that a grain needs to become net accelerated in its subsequent motion is larger than the largest possible lift-off velocity of splash-ejected grains \citep{Pahtzetal21}.

The velocity scale $\sqrt{\tilde gd}$, associated with splash ejection, dictates the scaling of the average streamwise velocity of saltating grains in the widely-used \cite{UngarHaff87}-scaling (Eq.~(\ref{UH87})) of the saturated saltation mass flux $Q$ \citep{Valanceetal15}. Our result that $\sqrt{\tilde gd}$ has no significant effect on the average saltation kinematics therefore represents a strong argument against this scaling and in favor of the \cite{Duranetal11}-scaling (Eq.~(\ref{D11})) of $Q$, the latter of which only requires considering the rebound dynamics of saltating grains, but not splash ejection \citep{PahtzDuran20}. This in turn supports the recent controversial claim that the scaling of $Q$ is insensitive to soil cohesion \citep{Comolaetal19a,Pahtzetal21}, since this rebound dynamics, in contrast to splash ejection, is not much affected by cohesive bonds between bed grains \citep{Pahtzetal21}.

Our results seem to be at odds with the recent wind tunnel experiments by \cite{Andreottietal21}, who measured that the wavelength $\lambda$ of aeolian impact ripples emerging from a flat bed is insensitive to the atmospheric conditions across a large range of the particle-fluid-density ratio ($s\in(2.1\times10^3,1.3\times10^6)$). In fact, these authors were able to theoretically explain their measurements only when assuming the \cite{UngarHaff87}-scaling of $Q$, whereas the \cite{Duranetal11}-scaling would not have worked for them, contrary to our main conclusion here. A possible explanation for this discrepancy lies in the fact that, for most of their tested atmospheric conditions, sand transport was undersaturated, possibly strongly undersaturated, at the end of the wind tunnel test section and that this undersaturation was mainly because most grains did not have nearly enough fetch to be net accelerated toward their steady state velocity \citep{Andreottietal21}. Under such circumstances, it seems, indeed, reasonable that most saltating grains exhibit a velocity close to the one associated with their splash ejection from the bed, that is, $\propto\sqrt{\tilde gd}$. In other words, the insensitivity of $\lambda$ to the atmospheric conditions measured by \cite{Andreottietal21} may, indeed, be explained using the \cite{UngarHaff87}-scaling. However, one should be cautious with extrapolating this finding to field conditions on Mars, where sand transport may have had a sufficiently long fetch to saturate during the initial phase of ripple formation. In fact, the \cite{Duranetal11}-scaling, favored by our study for saturated conditions, leads to a fundamentally different dependency of $\lambda$ on the atmospheric conditions \citep{Duranetal14b}.

Apart from weighing in on the ongoing debate about the scaling of $Q$, our results may also improve predictions of aeolian dune formation. In fact, the wavelength of dunes emerging from a flat bed scales with the so-called \textit{saturation length} $L_s$ \citep{Sauermannetal01,Kroyetal02b}, the scaling of which is controversial \citep{ClaudinAndreotti06,Andreottietal10,Pahtzetal13,Pahtzetal15b,LammelKroy17,JenkinsValance18,Selmanietal18}. A reasonable assumption is that $L_s$ scales with the saltation length $L$ above the aerodynamic entrainment threshold \citep{LammelKroy17}. If this assumption turns out to be true, our study will provide the simple prediction $L_s\propto\nu^{2/3}\tilde g^{-1/3}$ across a large range of atmospheric conditions, including typical conditions on Mars.

\section*{Acknowledgments}
The experimental data shown in the figures of this article can be found in the following references: \cite{Rasmussenetal96}, \cite{Andreotti04}, and \cite{Hoetal14}. We acknowledge support from grant Young Scientific Innovation Research Project of Zhejiang University (529001*17221012108).

%\bibliography{model}

\end{document}